\documentclass[11pt,a4paper]{article}
\usepackage[hyperref]{emnlp2018}
\usepackage{times}
\usepackage{latexsym}

\usepackage{url}

\usepackage{amsmath,amssymb,amsfonts}
\usepackage{algorithmic}
\usepackage{graphicx}
\usepackage{textcomp}
\usepackage{xcolor, colortbl}
\usepackage{multirow}
\usepackage{array}
\usepackage{subcaption}

\usepackage{array}
\newcolumntype{L}{>{\centering\arraybackslash}m{2.75cm}}
\usepackage{enumitem}
\setlist[enumerate]{leftmargin=*}
\setlist[itemize]{leftmargin=*}

\aclfinalcopy 

\title{Creative Procedural-Knowledge Extraction From Web Design Tutorials}

\author{
Longqi Yang$^1$ $\ \quad$ Chen Fang$^2$ $\ \quad$ Hailin Jin$^3$ $\ \quad$ Walter Chang$^3$ $\ \quad$ Deborah Estrin$^1$\\[0.1em]
$^1$Cornell Tech, Cornell University $\quad$ $^2$ByteDance AI Lab $\quad$ $^3$Adobe Research\\[0.1em]
\texttt{$^1$\{ly283, destrin\}@cornell.edu} $\quad$ \texttt{$^2$fangchen@bytedance.com}\\[0.1em]
\texttt{$^3$\{hljin, wachang\}@adobe.com}
}

\date{}

\begin{document}
\maketitle
\begin{abstract}
  Complex design tasks often require performing diverse actions in a specific order. To (semi-)autonomously accomplish these tasks, applications need to understand and learn a wide range of design procedures, i.e., Creative Procedural-Knowledge (CPK). Prior knowledge base construction and mining have not typically addressed the creative fields, such as design and arts. In this paper, we formalize an ontology of CPK using five components: \textit{goal}, \textit{workflow}, \textit{action}, \textit{command} and \textit{usage}; and extract components' values from online design tutorials. We scraped 19.6K tutorial-related webpages and built a web application for professional designers to identify and summarize CPK components. The annotated dataset consists of 819 unique commands, 47,491 actions, and 2,022 workflows and goals. Based on this dataset, we propose a general CPK extraction pipeline and demonstrate that  existing text classification and sequence-to-sequence models are limited in identifying, predicting and summarizing complex operations described in heterogeneous styles. Through quantitative and qualitative error analysis, we discuss CPK extraction challenges that need to be addressed by future research.
\end{abstract}

\section{Introduction}

Building applications that can enhance human abilities to accomplish creative tasks (such as writing summaries, designing graphics, and composing music) has recently captured a significant amount of attention from academia and industry~\cite{Ganin2018SynthesizingPF, performance-rnn-2017, See2017GetTT}. In the domain of design and imaging, these tasks 
often require micro-controls over pixels, shapes, and objects and involve multiple steps of operations. For example, \textit{drawing a creative cartoon} typically requires several iterations of an action sequence (contouring, brushing, coloring) applied to different areas on the canvas. Therefore, to realize a wide range of intelligent tasks (such as next-tool recommendation, auto-completion, personalized interface, and design material retrieval), applications need to gain \textbf{C}reative \textbf{P}rocedural-\textbf{K}nowledge (\textbf{CPK}).



In this paper we exploit the opportunity for machine learning algorithms to extract CPK from the growing corpus of online tutorials. Creative professionals document detailed actions and steps for using design software (e.g., Photoshop, GIMP, Inkspace, and Autodesk) in online design tutorials. These tutorials can be in the form of text, image, or video and may vary in quality, length, and topics. For example, \textit{envatotuts+} website\footnote{https://design.tutsplus.com} contains tutorials for text effects, sketches, watercolor painting, and animation. To the best of our knowledge, few prior work investigated the problem of parsing and extracting knowledge from free-text tutorials. Existing research on mining web content has been mostly focused on articles~\cite{ren2017cotype} and scientific papers~\cite{zhang2015deepdive}, and the tasks are limited to tagging~\cite{joulin2017bag}, name entity recognition~\cite{manning2014stanford}, and summarization~\cite{See2017GetTT}. Prior work on parsing cooking recipes~\cite{jermsurawong2015predicting, chen2017statistical} investigated clean and structured documents where ingredients are presented upfront, and instructions are described step-by-step without irrelevant or redundant information. In comparison, tutorials are much more diverse and unstructured, and the CPK extends beyond name entities, topics, and a fixed list of ingredients.

\begin{figure}
	\centering
	\includegraphics[width=1.0\columnwidth]{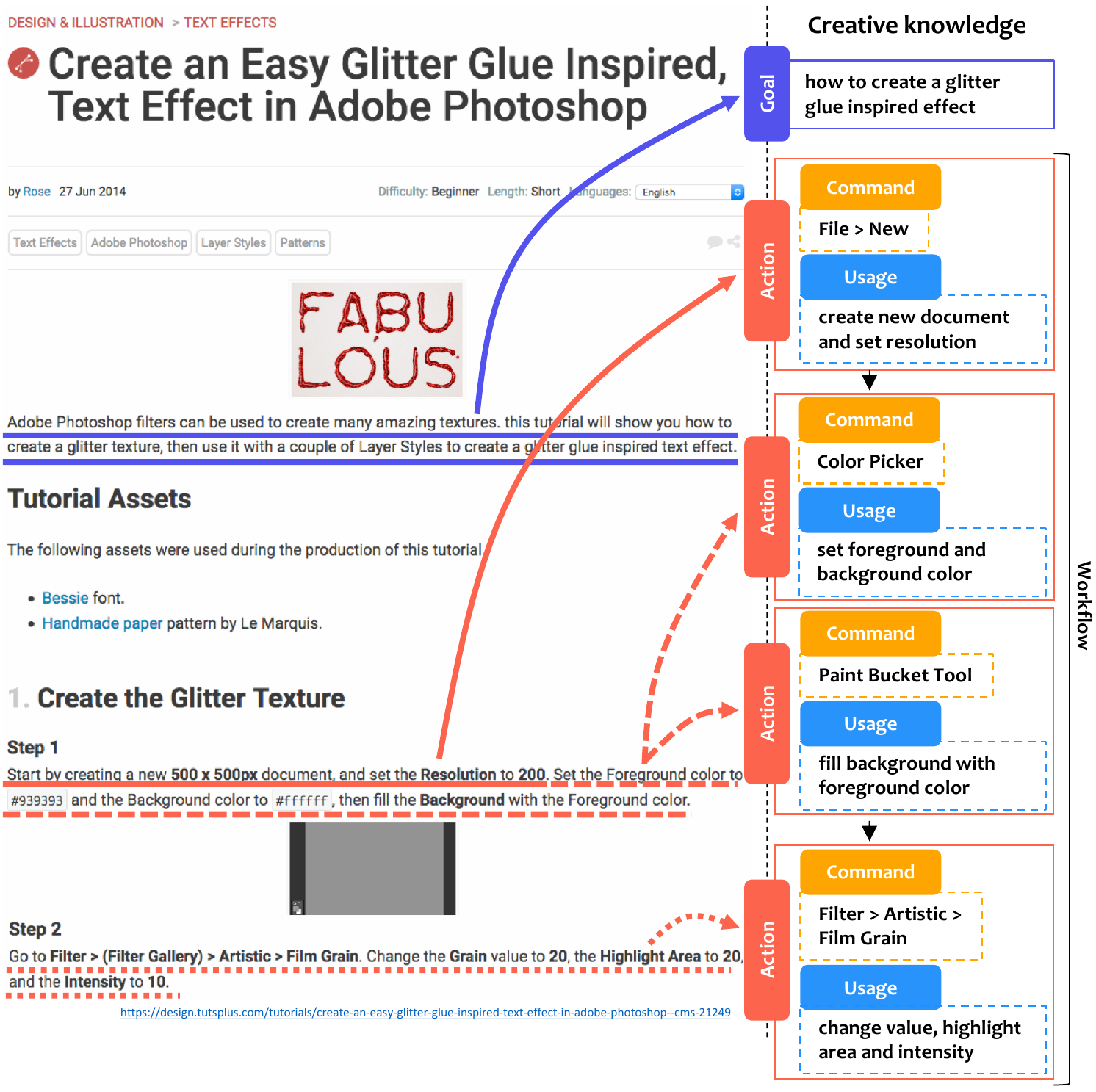}
	\caption{Sample CPK (Section \ref{sec:ontology}) annotations for a design tutorial. A professional designer is instructed to identity relevant text chunks and summarize  for the goal and actions. For each action, a command is also labeled.}
	\label{fig:annotation_example}
	\vspace{-5mm}
\end{figure}

To facilitate and benchmark research, we propose an ontology for CPK using the workflow representation that consists of \textit{goal}, \textit{workflow}, \textit{action}, \textit{command} and \textit{usage} (Section \ref{sec:ontology}). Based on the defined ontology, we collect thousands of text-based Adobe Photoshop tutorials from hundreds of domains and build a web application to elicit labels from professional designers. Our application allows annotators to flexibly extract and summarize various components and specify the sources from which the knowledge is derived. Finally, our dataset contains 819 unique commands, 47,491 actions, and 2,022 workflows and goals. A representative set of annotations is shown in Fig.~\ref{fig:annotation_example}. The code and dataset are available at \url{https://github.com/ylongqi/cpk}. With the dataset, we quantitatively and qualitatively present the challenges of CPK extraction. Specifically, we demonstrate that existing text classification and sequence-to-sequence models~\cite{sutskever2014sequence} fall short in handling heterogeneous writing styles of different tutorials. As a result, predictive models are suboptimal in identifying and classifying implicit- or latent-mentioned commands and produce low quality summaries for complex operations.

Since CPK stores valuable procedural knowledge necessary for diverse design tasks, it can potentially enable or enhance a wide range applications. For example, (a) \textbf{smarter tutorial search and recommendation.} CPK can enable command, action and intention-based search that are beyond text matching. Similarly, tutorial recommendations can be conducted based on the goal that an user plans to achieve or the commands and operations that she aspires to learn; (b) \textbf{skill trees and personalized curriculum.} A large collection of CPK provides resources to contruct skill trees that encode the \textit{rank} of each operation, which can power personalized curriculum, and (c) \textbf{autonomous design agents.} CPK provides mappings from design goals to action and command sequences. An autonomous design agent can be trained supervisedly to act on users' queries; In addition, since each annotated CPK directly corresponds to a tutorial, the collected dataset can be used for tutorial generations based on given command sequences and goal descriptions.

\section{Related Work}
Our work is related to and inspired by previous research on knowledge base construction and document modeling. The extracted CPK has applications to computational creativity.

\subsection{Knowledge base construction}
A knowledge base (graph)~\cite{bollacker2008freebase, liu2010biosnowball} stores facts using triplets ($e_i$, $e_j$, $r$), where $e_i$ and $e_j$ are generic entities (e.g., person, location, and event), or domain-specific entities (e.g. drug and effect), and $r$ is a relation that builds connections between $e_i$ and $e_j$ (e.g., a drug has a effect, and a person lives in a location). The process of Knowledge Base Construction (KBC)~\cite{niu2012elementary, mitchell2018never, mahdisoltani2013yago3, ritter2013extracting, zhang2015deepdive} is to extract triplets from free text~\cite{mitchell2018never, liu2017heterogeneous} or structured tables~\cite{ran2015domain, crestan2010web}. Prior work has proposed diverse approaches for the extraction, such as supervised methods~\cite{bunescu2005shortest}, semi- or weakly supervised algorithms~\cite{liu2010biosnowball, mahdisoltani2013yago3, mitchell2018never, jiang2012learning}, and distantly-supervised algorithms~\cite{liu2017heterogeneous, zhang2015deepdive, surdeanu2012multi, angeli2014combining, ren2017cotype}. These approaches require different levels human labeling efforts and are shown to vary in precision and recall. However, while existing knowledge bases have enabled important applications (such as web search~\cite{deshpande2013building} and question answering~\cite{yih2016question}), the triplet representations are over-simplified for design and creative tasks. These tasks often involve complex workflows, which require sequential and nested structures beyond binary relationships between entities. Our work compensates existing KBC literature by proposing a new ontology specifically tailored for CPK. We demonstrate that CPK extraction can benefit from and pose new challenges to existing approaches.

\subsection{Document modeling}
Existing work on document modeling has mainly focused on word-level and document(sentence)-level predictive tasks, e.g., name entity recognition, part-or-speech tagging, and semantic parsing~\cite{manning2014stanford} are used to tag word-level structures; and document representations~\cite{Le2014DistributedRO, Lau2016AnEE} are leveraged for sentiment analysis~\cite{Maas2011LearningWV}, textual similarity comparison~\cite{Maas2011LearningWV}, question retrieval~\cite{Hoogeveen2015CQADupStackAB}, and summarization~\cite{See2017GetTT}. Since many end applications are built on global contexts, it is non-trivial to extract documents' internal structures using trained predictive models, even with the attention mechanism~\cite{bahdanau2014neural, See2017GetTT}. Prior work on parsing cooking recipes~\cite{jermsurawong2015predicting, chen2017statistical} produced structured outputs on template documents, i.e., ingredients are known, and no redundant information is presented. In our work, we explicitly mine goals and design procedures from free-text tutorials, which are much more complex and unstructured. Such an information extraction poses challenges to existing modeling frameworks for word or sentence-level tasks and clean documents. But extracted fine-grained procedures may benefit traditional document classification, retrieval and summarization tasks.

\subsection{Computational creativity}
Recently, the field of computational creativity~\cite{Colton2012ComputationalCT} has captured growing attention from academia and industry. The goal of this field is to create programs that can master or enhance human-level creativity\footnote{https://en.wikipedia.org/wiki/Computational\_creativity}. In the domain of design and arts, researchers have proposed algorithms that can color grey images~\cite{Zhang2016ColorfulIC}, transfer image contextual styles~\cite{Li2017UniversalST}, and synthesis images~\cite{Sangkloy2017ScribblerCD, Ganin2018SynthesizingPF}. In addition, many applications, such as command search~\cite{Adar2014CommandSpaceMT}, command recommendation~\cite{Li2011DesignAE}, and creative content recommendation~\cite{Yang2017PersonalizingSA}, have been designed to assist creative professionals to accomplish complex tasks. Moving forward, to truely understand creativity, it is important to learn fundamental procedures of design. To that end, our exploration provides initial resources and insights for scalable discovery and learning of CPK in the future.

\section{Ontology of creative procedural-knowledge}
\label{sec:ontology}

\begin{figure}
	\centering
	\includegraphics[width=0.9\columnwidth]{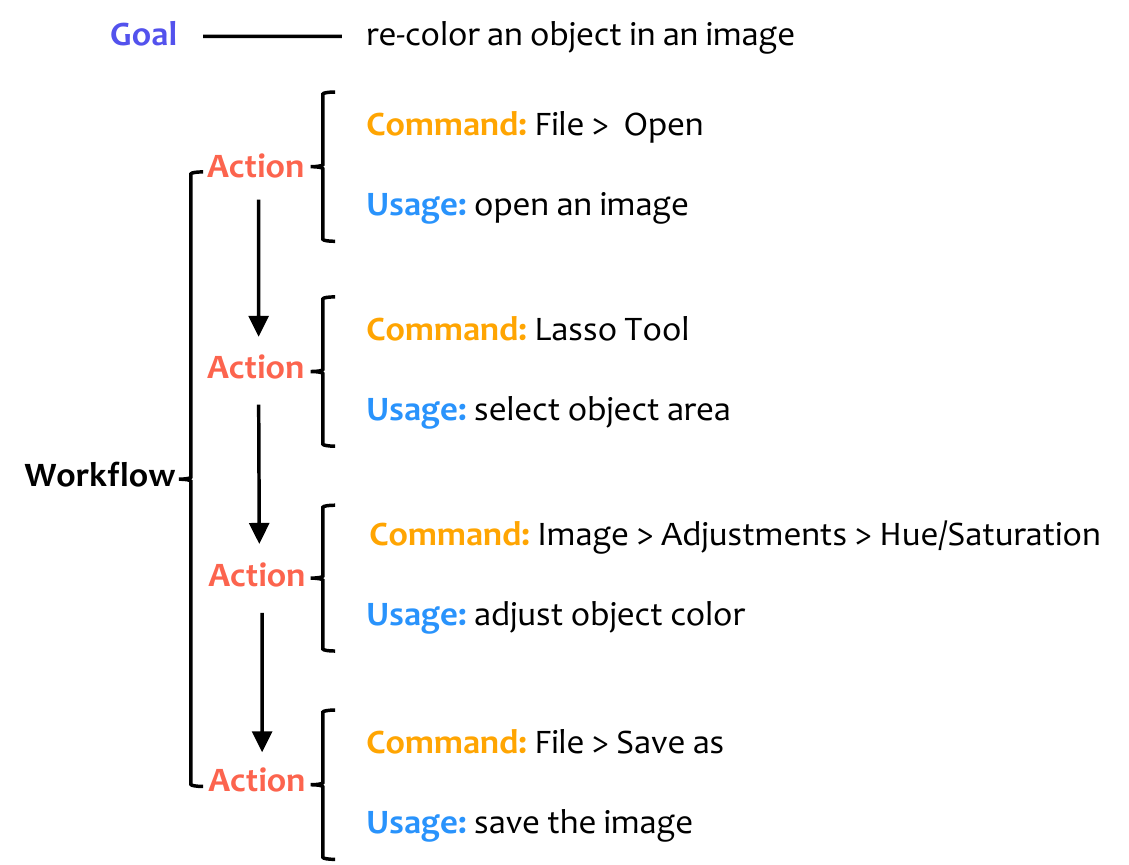}
	\caption{A CPK ontology for \textbf{the object re-coloring task}. CPK components are colored and bolded.}
	\label{fig:sample_ontology}
	\vspace{-5mm}
\end{figure}

Each design using software is fundamentally a \textit{process} that achieves an artistic objective by completing a sequence of actions, e.g., to re-color an object in an image using Photoshop, one needs to (1) load a target image into the software, (2) select object area, (3) adjust color, and (4) save the modified image. All processes that designers developed for diverse tasks form the CPK. To encode process-oriented nature, we compose an ontology of CPK using the \textit{workflow representation}, which was adopted to characterize business and scientific processes~\cite{bergmann2014similarity}. In principle, we represent each design process with two major components, a \textbf{goal} and a \textbf{workflow}, and the workflow ensembles \textbf{actions} (i.e., (\textbf{command}, \textbf{usage}) tuples) in a structured manner. A concrete implementation of the ontology for the \textit{object re-coloring} task is shown in Fig.~\ref{fig:sample_ontology}. Below are detailed definitions of \textit{goal} and \textit{workflow}.

\begin{itemize}
	\item \textbf{Goal.} The goal defines the objective for a design process. It typically describes a targeted artifact or a visual effect. For example, in terms of digital painting, the goal can be ``create character concept art'', ``paint water, waves and ocean'', or ``turn a pencil sketch into a colorful character illustration''; and in terms of photo effects, potential goals are ``create an architecture sketch effect'', ``add lights to a tree'', or ``make someone look younger''. In general, goals are not mutually exclusive, and they more or less relate to each other, e.g., achieving higher-level goals may depend on the completion of lower-level tasks (inclusion), and an abstract goal may correspond to multiple concrete implementations (hierarchy). In Fig.~\ref{fig:sample_ontology} example, the goal is to ``re-color an object in an image''.
	
	\item \textbf{Workflow.} The workflow represents unrolled and step-by-step actions that need to be performed to accomplish the goal. Each action specifies a software \textbf{command} to be used and its \textbf{usage} (i.e., what it is used for). For example, the workflow for the task in Fig.~\ref{fig:sample_ontology} contains four actions (``\{\}'' and ``[]'' represent an usage and a command respectively): (1) use [File $>$ Open] to \{open an image\}, (2) employ [Lasso Tool] to \{select object area\}, (3) \{adjust object color\} through [Image $>$ Adjustments $>$ Hue/Saturation], and (4) \{save the image\} using [File $>$ Save as]. Similar to the goal definition, workflows also correlate with each other, e.g, two workflows may share a similar sub-action sequence. In reality, a comprehensively defined workflow should be executable given an environmental context (i.e., an image).
\end{itemize}

The canonical knowledge representation makes it possible to annotate and extract CPK from online design tutorials.

\section{Annotating Creative Procedural-Knowledge}

\begin{figure}
	\centering
	\includegraphics[width=0.9\columnwidth]{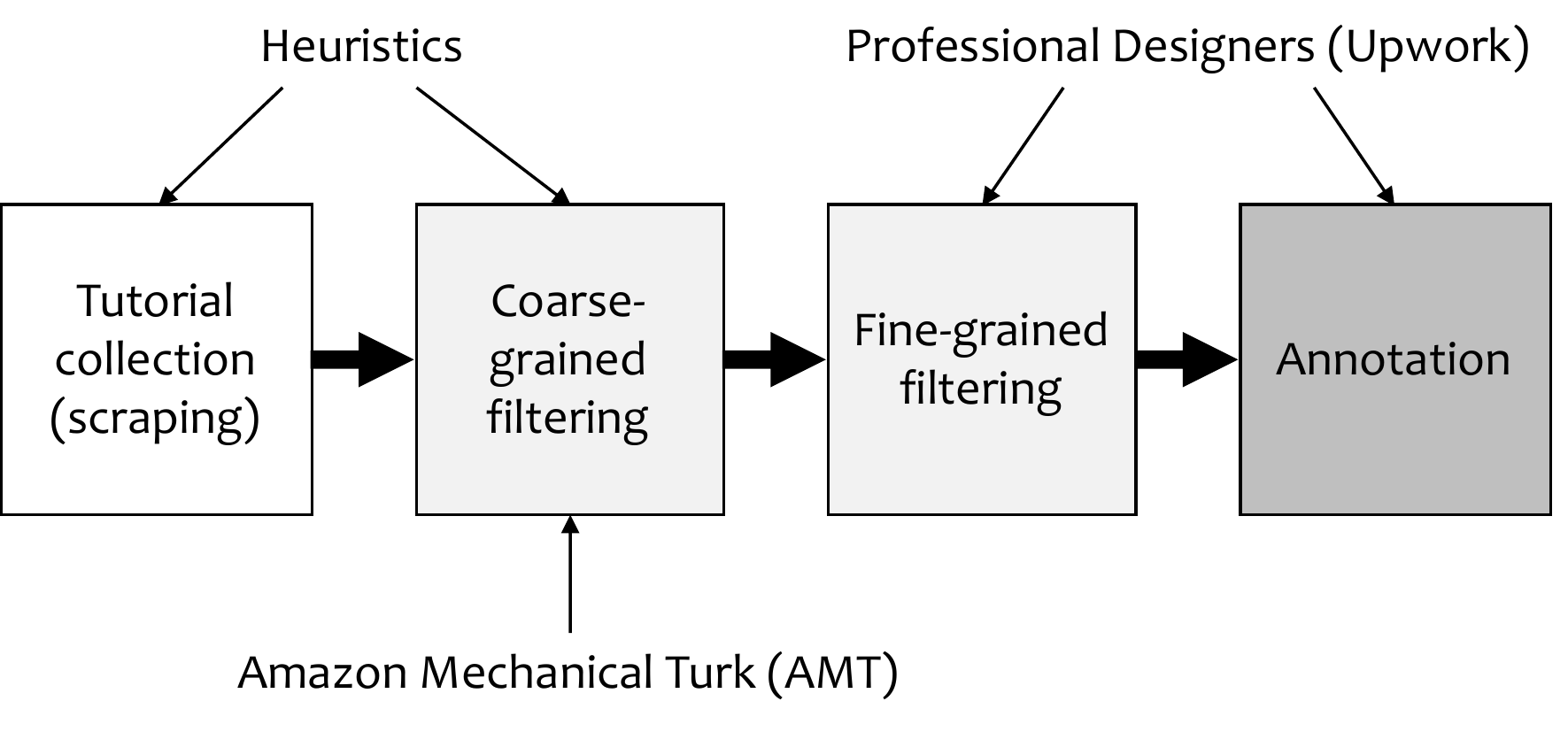}
	\caption{The CPK annotation pipeline. It contains three steps: scraping, filtering (coarse- and fine-grained) and annotation.}
	\label{fig:ann_pipeline}
\end{figure}

We collect CPK from design tutorials (Fig.~\ref{fig:annotation_example}) that explicitly or implicitly describe the task that it tries to accomplish and detailed step-by-step instructions. To capture a wide range of goals and workflows, we consider Adobe Photoshop as the design software, because of its rich functionality and active user community. In this section, we describe  processes of collecting, filtering and annotating tutorial webpages. As shown in Fig.~\ref{fig:ann_pipeline}, the filtering contains two stages, coarse-grained and fine-grained, that are conducted through Amazon Mechanical Turk (AMT) and professional designers from Upwork, respectively.

\subsection{Tutorial collection}

Using Google search engine, we identify 148 valid and unique domains, where each domain serves more than 10 Photoshop tutorials. We build a generic web crawler to recursively scrape all web pages under these domains and only retain pages with the keyword ``photoshop''. To further improve the accuracy of the scraping, for the 27 largest domains, we build dedicated scrapers tailored for their website structures. Eventually, 19.6K web pages are collected. We use Simhash~\cite{manku2007detecting} to detect duplicated documents, which results in 18,100 distinct pages for further processing.

\subsection{Course-grained filtering}

\begin{figure}
	\centering
	\includegraphics[width=0.9\columnwidth]{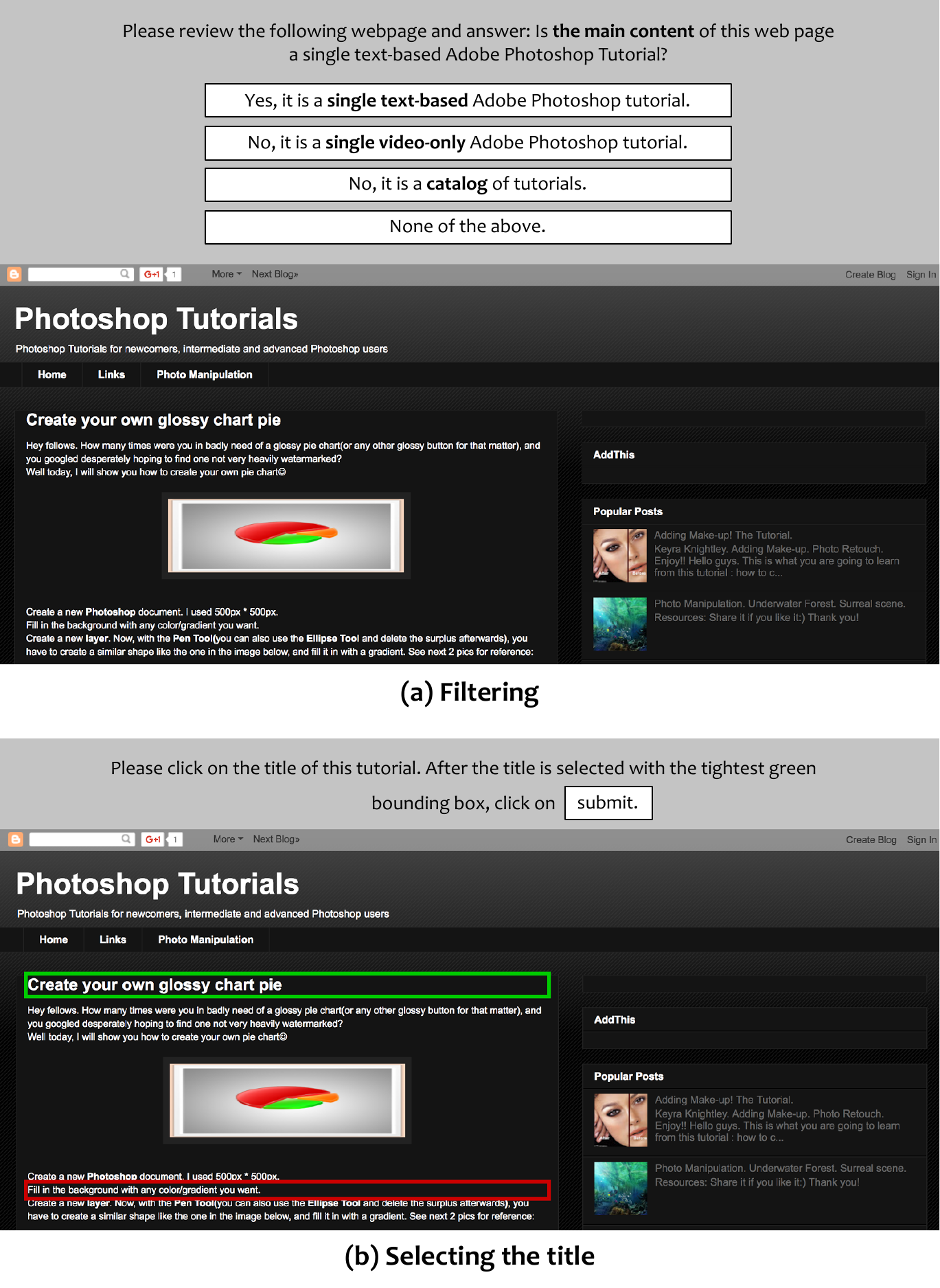}
	\caption{The web application for coarse-grained webpage filtering on Amazon Mechanical Turk (AMT) platform. Each worker is instructed to complete two tasks: answer a filtering-related question, and select the title of the tutorial.}
	\label{fig:amt_filtering}
	\vspace{-3mm}
\end{figure}

The scraped web pages are potentially noisy. For example, pages may not be Photoshop tutorials. Also, since we mainly focus on text-based tutorials, web pages that only contain videos are out of our scope. Because identifying these characteristics does not require professional design knowledge, we conduct a coarse-grained filtering using Amazon Mechanical Turk (AMT) platform. For each web page, two distinct workers are recruited to use a web application (Fig.~\ref{fig:amt_filtering}) to (a) answer the question ``Is the main content of this web page a single text-based Adobe Photoshop Tutorial?'', and (b) click on the title of this tutorial (if the answer is ``Yes'' to the previous question). The second task is designed to verify the simple click made for the first question. Qualified workers need to satisfy three requirements: (1) ``masters'' as determined by AMT platform, (2) have more than 90\% of approval rate, and (3) have completed more than 100 tasks. The workers are paid for \$10/hour. Finally, 9996 pages receive consistent ratings from both workers.

\subsection{Fine-grained filtering and annotation}

\begin{figure*}
	\centering
	\includegraphics[width=2.0\columnwidth]{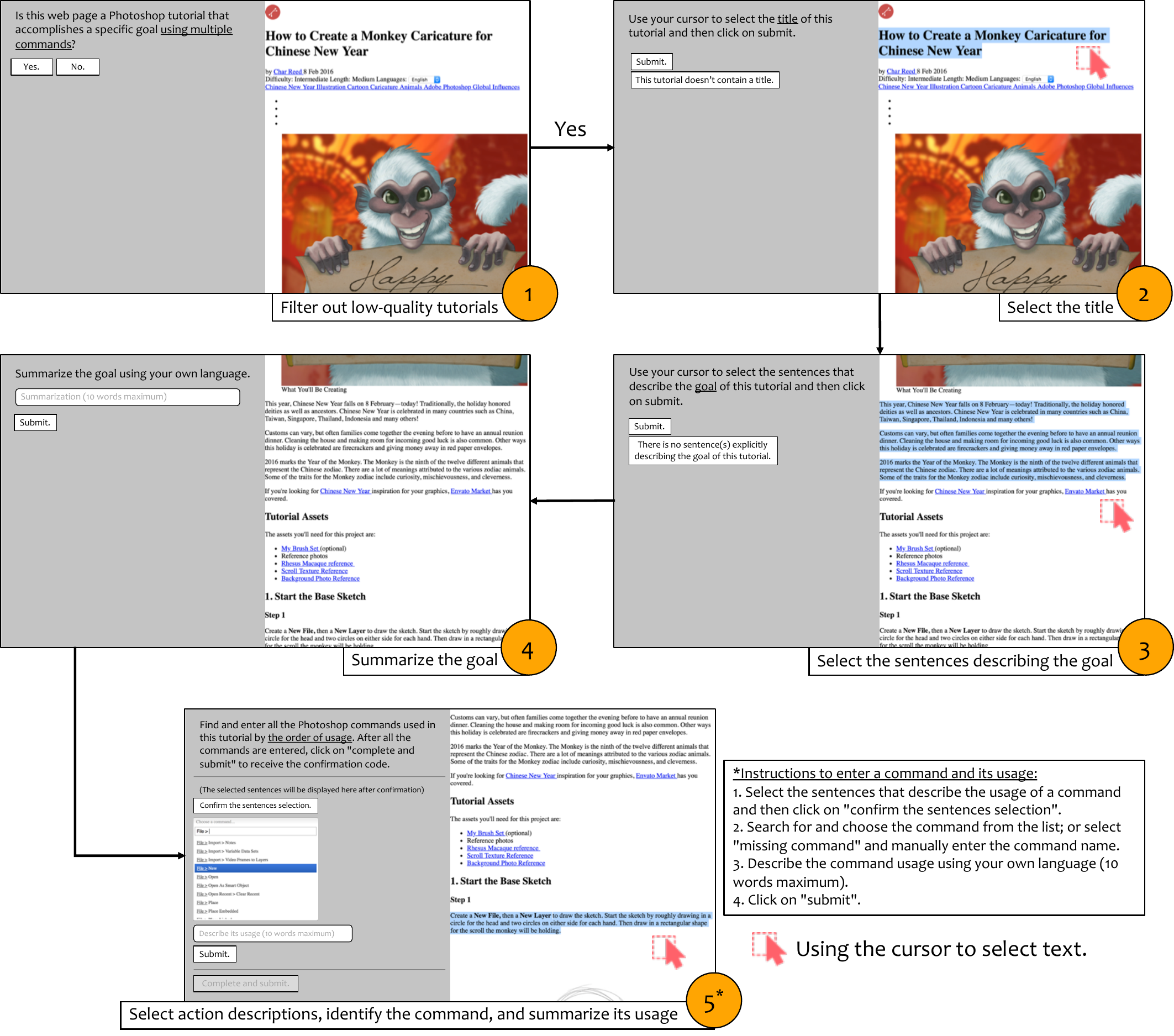}
	\caption{The web application for professional designers to conduct fine-grained filtering and CPK annotations. Each session consists of five steps: (1) filter out low-quality tutorials, (2) select the title of the tutorial, (3) select the sentences describing the goal, (4) summarize the goal using less than 10 words, (5) annotate actions in the order of their usage (continue until all actions are identified). The text can be directly selected using the cursor.}
	\label{fig:designer_filtering}
\end{figure*}

Cleaned tutorial pages vary in quality, e.g., some pages may only contain simple tips, while others contain detailed steps for complex tasks. To achieve quality consistency, we recruit 8 professional designers from Upwork platform to filter out low-quality tutorials (those have unclear goals or use few commands) and annotate high-quality ones using a web application (Fig.~\ref{fig:designer_filtering}). Recruited designers have 2-10 years of Photoshop experience and are fluent in English. For each tutorial page, a designer is assigned to (a) select the text that describes the title and the goal, (b) use natural language to summarize the goal (with less than 10 words), and (c) identify actions performed. Annotations for an action contain the original text, an used command, and summarization (10 words maximum) for its usage. These annotations map unstructured design tutorials into structured CPK (Section \ref{sec:ontology}). 

\subsection{Dataset overview}

\begin{figure}
	\centering
	\includegraphics[width=1.0\columnwidth]{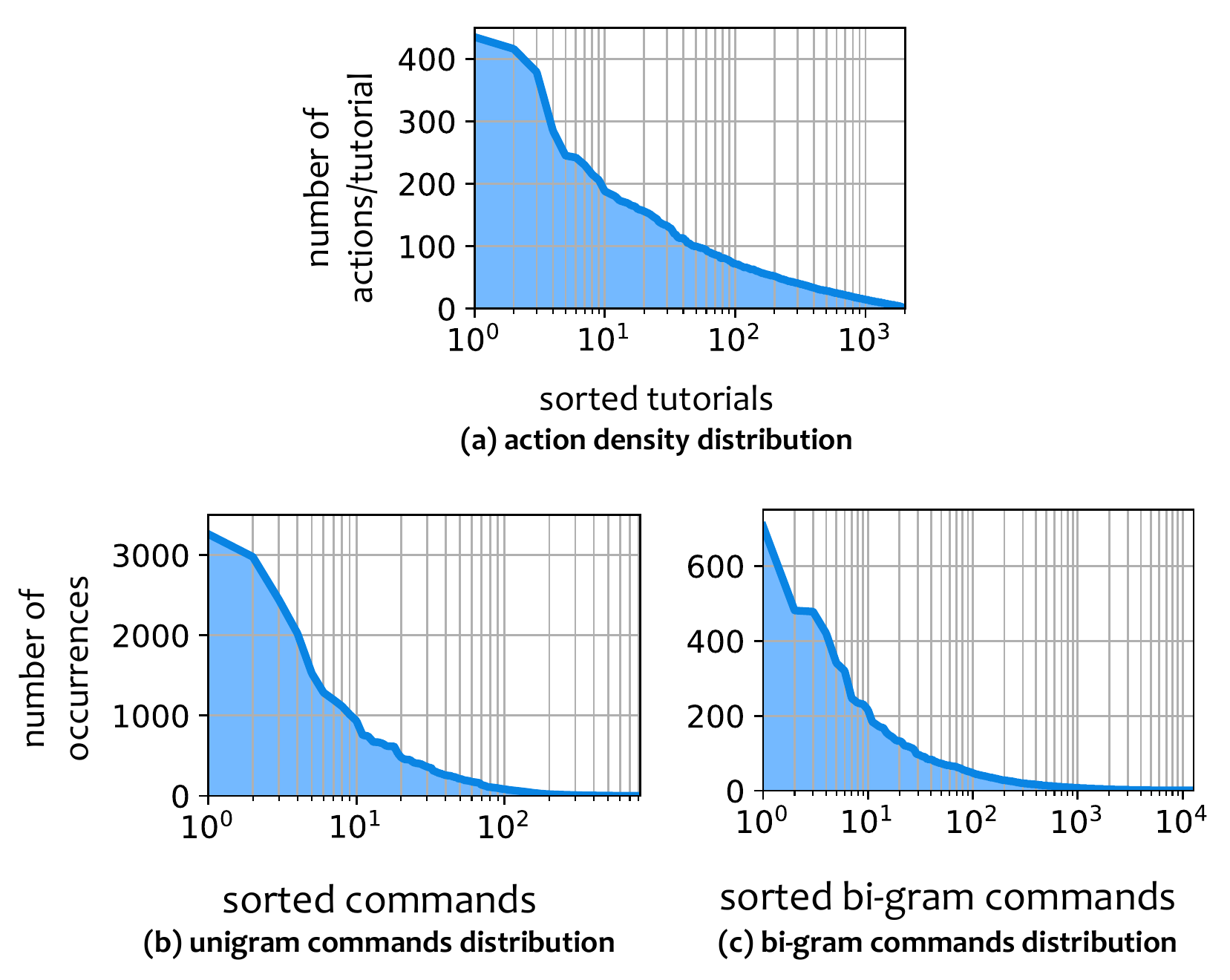}
	\caption{Action and command distributions in the annotated CPK dataset. X-axises are log-scaled for better visualization.}
	\label{fig:dist}
\end{figure}


\begin{table}[]
	\centering
	
	\begin{tabular}{c|c} \hline
		\textbf{unigram} & \textbf{bigram} \\ \hline
		Brush Tool & New Layer*$\rightarrow$Brush Tool \\
		New Layer* & Blend Mode$\dagger$$\rightarrow$Opacity$\dagger$ \\
		Blend Mode$\dagger$ & Brush Tool$\rightarrow$New Layer* \\
		Duplicate Layer$\ddagger$ & Edit$>$Copy$\rightarrow$Edit$>$Paste \\
		File$>$Open & Brush Tool$\rightarrow$Brush Tool \\ \hline
	\end{tabular}
	\caption{Five most frequently annotated unigram and bigram commands in the dataset (*:Layers, $\dagger$: Layer Panel, $\ddagger$: Layer) \label{tbl:top_commands}}
	\vspace{-3mm}
\end{table}

After completing annotations, 819 unique commands and 47,491 actions in 2,022 tutorials are identified. We show the long-tail distribution of action density (number of actions per tutorial) in Fig.~\ref{fig:dist}-(a) --- Around 90\% of tutorials have less than 50 actions. In terms of the commands used, their frequencies are also skewed, as shown in Fig.~\ref{fig:dist}-(b), (c). The most frequently used tools and their combinations (Table.~\ref{tbl:top_commands}), such as Brush Tool, relate to common tasks in digital designs, e.g., manipulate layers and areas. These demonstrate that many complex design tasks do not necessarily require rare tools. 

For the goal and usage, the annotations are in the form of free text. We find that designers tend to use boilerplate phrases to summarize the goal, such as ``how to'', ``learn to'', and ``learn how to''. We remove these phrases using regular expressions since they do not provide content information. To visualize the goal summaries, we first derive summary representations using the element-wise average of GloVe word vectors~\cite{pennington2014glove} (excluding stop words), and then apply the K-means algorithm to discover underlying clusters. We set $K=5$, and the summaries closest to each cluster center are presented in Table.~\ref{tbl:goal_k_means}. The five clusters cover a wide range of design topics, e.g., text effect, photo effect, web design, textures, and scenes, which demonstrates the diversity of the dataset. In addition, we apply the same clustering approach to the usage summaries of all commands. As a result, commands are grouped by their usage similarities. For example, as shown in Table.~\ref{tbl:tools_clusters}, filter- and transform-related commands are grouped into the cluster 1 and cluster 5, respectively. This demonstrates that CPK reveals semantic relationships between different commands.

\begin{table*}
\fontsize{7.5}{12.2}\selectfont
	\centering
	\begin{tabular}{L|L|L|L|L}\hline
		\rowcolor{gray}
		\textbf{Cluster 1} & \textbf{Cluster 2} & \textbf{Cluster 3} & \textbf{Cluster 4} & \textbf{Cluster 5}\\ \hline
		create photo \textcolor{gray}{having} pretty good light effects & create dreamy photo effect using simple tools \textcolor{gray}{and} filters & create fun designs comning text \textcolor{gray}{and} shapes & create abstract photo using effects \textcolor{gray}{and} color & create \textcolor{gray}{this} trasparent text effect\\ \hline
		create like \textcolor{gray}{in} \textcolor{gray}{my} dream photo effect & create abstract photo using effects \textcolor{gray}{and} color & create picture \textcolor{gray}{by} blending different images \textcolor{gray}{and} tools & create \textcolor{gray}{a} painting \textcolor{gray}{from} \textcolor{gray}{a} photo & create \textcolor{gray}{a} text effect\\ \hline
		create realistic light around \textcolor{gray}{a} subject & creating \textcolor{gray}{a} cool matelic background using \textcolor{gray}{some} filters \textcolor{gray}{and} effect & create colorful futuristic looking text effect & create \textcolor{gray}{a} lomography photo effect & create \textcolor{gray}{a} text effect\\ \hline
		use fire \textcolor{gray}{and} blending modes \textcolor{gray}{to} create got theme photos & create \textcolor{gray}{a} fabric text effect using layer styles \textcolor{gray}{and} texture & creating \textcolor{gray}{a} landscape creative image using structure images & create a-smoke photo effect & create text effect\\ \hline
		create \textcolor{gray}{a} night scene image & creating \textcolor{gray}{a} text effect using texture \textcolor{gray}{and} filters & create \textcolor{gray}{a} modern web design style & create \textcolor{gray}{a} colorful photo manipulation & create \textcolor{gray}{a} brazil-inspired text effect \textcolor{gray}{in} \textcolor{gray}{photoshop}\\\hline
	\end{tabular}
	\caption{Five K-means clusters of goal summaries. The K-means algorithm is applied to summary representations, which are derived by taking the average of GloVe word vectors~\cite{pennington2014glove} excluding stop words (represented with \textcolor{gray}{light fonts}).\label{tbl:goal_k_means}}
\end{table*}

\begin{table*}
\fontsize{7.5}{12.2}\selectfont
	\centering
	\begin{tabular}{L|L|L|L|L}\hline
		\rowcolor{gray}
		\textbf{Cluster 1} & \textbf{Cluster 2} & \textbf{Cluster 3} & \textbf{Cluster 4} & \textbf{Cluster 5}\\ \hline
		Layer$>$New Adjustment Layer$>$Photo Filter & Pen Tool & Layer$>$New Adjustment Layer$>$Color Balance&Layers Panel&Edit$>$Transform$>$Warp \\ \hline
		Filter$>$Texture$>$Grain & Selection Tool & Layer$>$New Adjustment Layer$>$Selective Color & Eye Icon & Edit$>$Transform$>$Forward Warp \\ \hline
		Filter$>$Distort$>$Spherize & Polygonal Lasso Tool & Image$>$Adjustments$>$  Selective Color & Layers$>$Palette Options & Edit$>$Free Transform \\ \hline
		Filter$>$Texture$>$Texturizer & Move Tool & Image$>$Adjustments$>$ Variations & Layer$>$Layer Mask$>$Reveal All & Image$>$Transform$>$Free Transform \\ \hline
		Filter$>$Distort$>$Ripple & Rectangular Marquee Tool & Layer$>$New Adjustment Layer$>$Hue/Saturation & Layer$>$Layer Mask$>$Hide All & Edit$>$Transform$>$ Perspective \\ \hline
	\end{tabular}
	\caption{Five K-means clusters of commands. The K-means algorithm is applied to commands' usage summary representations, which are derived by taking the average of GloVe word vectors~\cite{pennington2014glove} excluding stop words.\label{tbl:tools_clusters}}
\end{table*}

\begin{table*}
\fontsize{7.5}{9.5}\selectfont
	\begin{subtable}{2.0\columnwidth}
		\centering
		\begin{tabular}{>{\centering\arraybackslash}m{0.2cm} |>{\centering\arraybackslash}m{8.5cm} | >{\centering\arraybackslash}m{2.4cm} | >{\centering\arraybackslash}m{2.4cm}}
			\hline
			\rowcolor{gray}\textbf{Id} & \textbf{Sentence} & \textbf{Prediction@1} & \textbf{Ground Truth} \\ \hline
			1 & make sure that you have a radial gradient that fades from solid white to transparent as shown in the image below : on your new layer , create a gradient and change the blending mode to overlay . & Gradient Tool & \begin{tabular}[c]{@{}c@{}}Gradient Tool\\ \hline Blend Mode\\ \hline New Layer\end{tabular} \\ \hline
			2 & copy ( ctrl + c ) and paste ( ctrl + v ) the selection . & Edit\textgreater{}Paste & \begin{tabular}[c]{@{}c@{}}Edit\textgreater{}Copy\\ \hline Edit\textgreater{}Paste\end{tabular} \\ \hline
			3 & create another new layer above , use it as clipping mask ( cmd/ctrl + alt + g ) again . & Create Clipping Mask & \begin{tabular}[c]{@{}c@{}}Create Clipping Mask\\ \hline New Layer\end{tabular} \\ \hline
			4 & step 2 right click the canvas and choose select inverse to invert the selection . &  Select\textgreater{}Inverse & \begin{tabular}[c]{@{}c@{}}Edit\textgreater{}Copy\\ \hline Edit\textgreater{}Paste\\ \hline Select\textgreater{}Inverse\end{tabular} \\ \hline
		\end{tabular}
		\caption{True Positive (TP) samples}
		\vspace{3mm}
	\end{subtable}
	
	\begin{subtable}{2.0\columnwidth}
		\centering
		\begin{tabular}{>{\centering\arraybackslash}m{0.2cm} | >{\centering\arraybackslash}m{14cm}}
			\hline
			\rowcolor{gray}\textbf{Id} & \textbf{Sentence} \\ \hline
			1 & next you will find a short version of the tutorial but for a more details please watch the video tutorial . \\ \hline
			2 & we want to sharpen objects in an image without increasing the effect or visibility of those unwanted elements and also in a way that does not affect the original colors . \\ \hline
			3 & i hope that you enjoyed this tutorial , and as always i'd love to hear what you think ! \\ \hline
			4 & my recommendation is to work in the early morning hours before the crowds set in . \\ \hline
		\end{tabular}
		\caption{True Negative (TN) samples (Prediction@1=Ground Truth=``No Action'')}
		\vspace{3mm}
	\end{subtable}
	
	\begin{subtable}{2.0\columnwidth}
		\centering
		\begin{tabular}{>{\centering\arraybackslash}m{0.2cm} |>{\centering\arraybackslash}m{8.5cm} | >{\centering\arraybackslash}m{2.3cm} | >{\centering\arraybackslash}m{2.3cm}}
			\hline
			\rowcolor{gray}\textbf{Id} & \textbf{Sentence} & \textbf{Prediction@1} & \textbf{Ground Truth} \\ \hline
			1 & then create a new shape using round shape tool . & Custom Shape Tool & Ellipse Tool \\ \hline
			2 & set it to soft light mode at 16 \% opacity . & Blend Mode & Gradient Map \\ \hline
			3 & ( i 've used \# 003200 here . ) & Paint Bucket Tool & Color Picker \\ \hline
			4 & then press ctrl+c ( win ) / command+c ( mac ) to copy the image to the clipboard . &  Edit\textgreater{}Copy & No Action \\ \hline
			5 & rename the merged layer to crackedplanet . & Rename Layer & No Action\\ \hline
		\end{tabular}
		\caption{False Positive (FP) samples}
		\vspace{3mm}
	\end{subtable}
	
	\begin{subtable}{2.0\columnwidth}
		\centering
		\begin{tabular}{>{\centering\arraybackslash}m{0.2cm} |>{\centering\arraybackslash}m{11cm} | >{\centering\arraybackslash}m{2.3cm}}
			\hline
			\rowcolor{gray}\textbf{Id} & \textbf{Sentence} & \textbf{Ground Truth}\\ \hline
			1 & step 21 : go to the background image layer at the bottom and press ctrl + j . & Duplicate Layer \\ \hline
			2 & painting it just a bit off could make her eyes look unparallel . & Clone Stamp Tool \\ \hline
			3 & draw two lines down from the corner of her eye , and tap it once or twice here and there : brushing the paint drip : zoomed out , this is what it looks like : time to fix her hair , we're going to plant a tree on top of her head , but we need her head to be a little more flat . & Brush Tool \\ \hline
			4 & adding bottom text step 1 follow the same techniques to add bottom text . & Horizontal Type Tool \\ \hline
			5 & hit command + d to deselect . &  Gradient Tool \\ \hline
			6 & once you have opened the actions panel you can begin creating your first lomo leak . & Actions Panel \\ \hline
		\end{tabular}
		\caption{False Negative (FN) samples (Prediction@1=``No Action'')}
		\vspace{3mm}
	\end{subtable}
	\caption{Commands predictions for sample sentences in the testing set. The samples are grouped into True Positive (TP), True Negative (TN), False Positive (FP), and False Negative (FN). The predictions are from the fastText model built on the 1,2,3,4-gram features. The Prediction@1 denotes the top label (either a command or a ``No Action'') predicted for each instance.\label{tbl:qua_command}}
\end{table*}

\section{Extracting creative procedural-knowledge}

The annotated dataset makes it possible to supervisedly build models to mine and extract CPK. In this section, we propose a general pipeline that produces structured knowledge outputs based on unstructured web content (Section \ref{sec:extraction_pipeline}). In addition, we experiment existing text classification and summarization algorithms for various components (Section \ref{sec:algorithm_experimentation}). We find and discuss limitations of existing approaches and challenges of the extraction task.

\subsection{A general extraction pipeline}
\label{sec:extraction_pipeline}

\begin{figure}
	\centering
	\includegraphics[width=1.0\columnwidth]{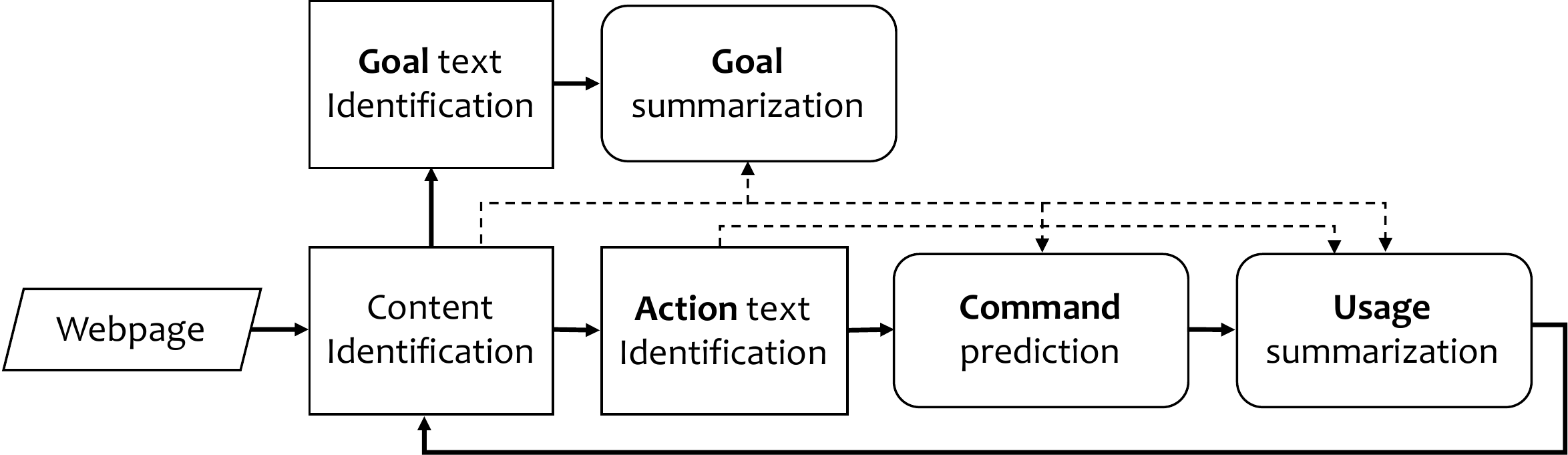}
	\caption{Proposed CPK extraction pipeline. All arrows represent data flows. Solid-line arrows additionally denote the order for executing modules. The parallelogram represents the input, Rectangulars represent identification tasks; and Rounded-rectangulars represent prediction/generation tasks.}
	\label{fig:extraction_pipeline}
\end{figure}

Inspired by the annotation procedure, we present an extraction pipeline in Fig.~\ref{fig:extraction_pipeline}. It consists of six steps: (1) extract main text content from a raw HTML tutorial webpage; (2) identify a text chunk that presents the goal of a design task; (3) summarize the goal; (4) identify text segments that describe actions recursively and orderly; (5) predict commands used for each action; and (6) summarize the usage of each command. The predictions and summarizations can leverage the local (identified text) and global (complete text content) information.

\subsection{Experimenting existing algorithms}
\label{sec:algorithm_experimentation}

The proposed pipeline decomposes the knowledge extraction task into several sub-tasks, which can be potentially powered by existing solutions for text analysis. For example, the command prediction can be viewed as a multi-label text classification problem and the usage summarization can leverage the sequence-to-sequence models~\cite{sutskever2014sequence} developed for machine translation. To explore the extent to which  existing solutions can solve the CPK extraction, we (a) apply the fastText~\cite{joulin2017bag} to identify action-related text chunks and predict commands, and (b) adapt the neural machine translation model~\cite{bahdanau2014neural} to summarize commands' usages. Specifically, we use the boilerpipe~\cite{kohlschutter2010boilerplate} and the newspaper\footnote{https://github.com/codelucas/newspaper} packages to extract main content from HTML pages, and assume the minimum unit for a text chunk is a sentence. As a result, an action may span across multiple sentences, and a sentence may contain multiple actions.

\subsubsection{Text chunks identification and commands prediction}
\label{sec:command_prediction}

\begin{figure}
	\centering
	\includegraphics[width=1.0\columnwidth]{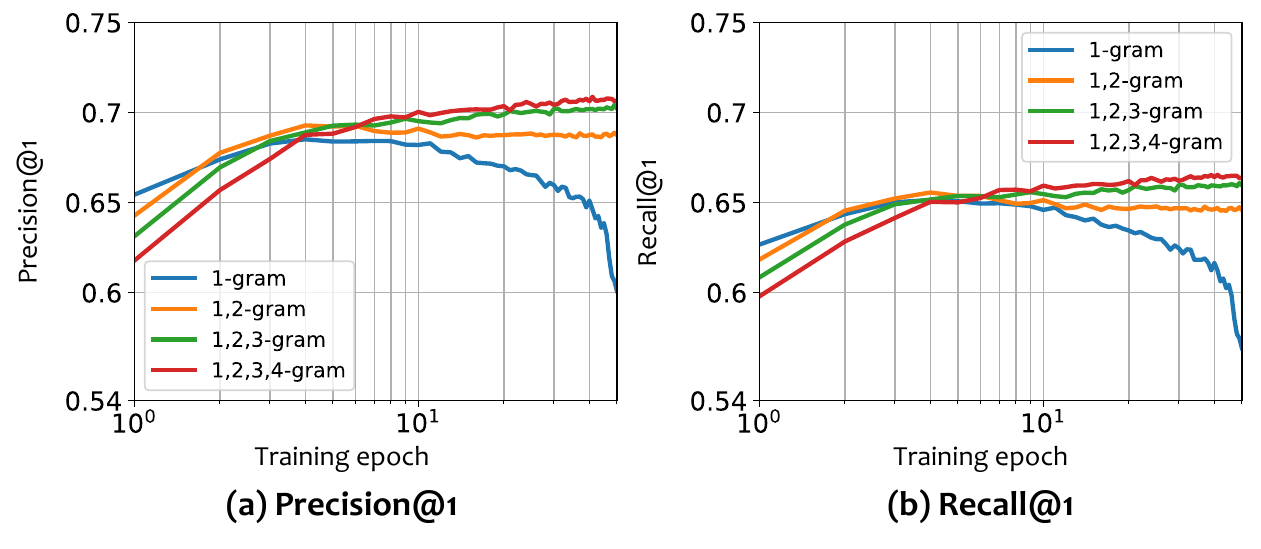}
	\caption{Precision and Recall for the command prediction task evaluated in the testing set. The performances of four fastText models built on different N-gram features are presented. }
	\label{fig:command_classification}
\end{figure}

To conduct sentence-level predictions, we use the NLTK~\cite{bird2009natural} to segment clean tutorial text into sentences. In total, 2022 tutorials are divided into 94,022 segments. For each sentence, a multi-label classifier predicts commands used or outputs ``No Action'' indicating that the sentence does not describe any action. In practice, the ``No Action'' is treated as an additional label, along with all commands that appear more than 5 times (404 out of 819 satisfy the requirement), in the classification. 54\% of sentences are ``No Action'' labeled. To measure the performance of fastText, we randomly split sentences into a training set (62,936 sentences) and a testing set (31,086 sentences), and experiment N-gram features, where N ranges from 1 to 4. The fastText classification performances, in terms of Precision@1 and Recall@1, are presented in Fig.~\ref{fig:command_classification}. Compared to the majority baseline (0.54), fastText performs significantly better, and the performance is further improved by adding more N-gram features. However, the improvements saturate when N is larger than 3, and the best performance is still unsatisfactory. 

To more deeply understand the characteristics of the task and its unique challenges, we analyze errors of the best classifier (built on the 1,2,3,4-gram features). Specifically, samples of True Positives (TP), True Negatives (TN), False Positives (FP), and False Negatives (FN) in the testing set are shown in Table.~\ref{tbl:qua_command} (``No Action'' is treated as negative). The TP samples demonstrate that commands predictions are accurate when they are explicitly mentioned in the text, e.g., ``create a gradient''(a-1), ``paste the selection''(a-2), ``use it as clipping mask'' (a-3), and ``select inverse to invert the selection'' (a-4), etc. Also, according to TN samples, ``No Action'' sentences are easy to classify when the semantics are clearly irrelevant to tutorial workflows, such as introductory and conclusive sentences (b-1 and b-3) and side notes (b-2 and b-4). However, as the sentences become more complex, the classifier fails under various scenarios, as discussed below.

\begin{itemize}
	\item \textbf{Implicit mention of commands.} The command names may not be explicitly mentioned in the text. Instead, they may be referred using short-cuts (d-1 and d-5) or appearances (c-1).
	\item \textbf{No direct mention of commands (Latent commands).} In many cases, commands are not mentioned at all. For example, in c-2, c-3, d-2, and d-3, text chunks describe the changes to be made to the canvas; and the usage of a command may be mentioned out of the given sentence, e.g., d-4 and d-5.
\end{itemize}

In order to tackle these challenges, we argue that prediction models need to (1) understand the global structure of a tutorial; and (2) develop deeper understandings of the relationships between commands and their applied consequences. Among the FP and FN samples, we find some labeling errors, such as c-4, c-5, and d-6. These errors may stem from the label mapping process, i.e., labels from an action are shared across sentences that the action spans, or are due to the mistakes made by annotators. We leave further data cleaning as future work.

\subsubsection{Usage summarization}

\begin{table}[]
\fontsize{7.5}{12.2}\selectfont
	\centering
	\begin{tabular}{c|ccc|ccc}
		\hline
		\multirow{2}{*}{dropout} & \multicolumn{3}{c|}{validation} & \multicolumn{3}{c}{testing} \\ \cline{2-7} 
		& 0 & 0.2 & 0.5 & 0 & 0.2 & 0.5 \\ \hline
		1-layer & 11.73 & 12.29 & 13.49 & 10.33 & 11.71 & 12.17 \\
		1-layer-att & 18.45 & 19.18 & \textbf{21.53} & 17.24 & 17.84 & \textbf{19.70} \\
		2-layer & 11.37 & 11.97 & 13.16 & 10.03 & 11.30 & 12.56 \\
		2-layer-att & 16.37 & 16.23 & 17.18 & 14.83 & 15.27 & 16.85 \\ \hline
	\end{tabular}
	\caption{BLEU scores for the usage summarization task evaluated in the validation and testing sets. The best performance in either set is bolded. The naming schema of different algorithms: [number of layers]-layer-[whether or not the attention mechanism~\cite{bahdanau2014neural} is applied].\label{nmt_results}}
	\vspace{-3mm}
\end{table}

An usage summarization module takes a raw sentence as the input and generates a command usage summary. A natural model that can accomplish this task is the sequence-to-sequence model~\cite{sutskever2014sequence}: a Recurrent Neural Network (RNN) based encoder ``reads'' a list of raw text tokens and produces a vector representation. Then a seperate RNN-based decoder takes the representation as the input and sequentially generates a list of words as the summary. We experiment the Neural Machine Translation (NMT) model~\cite{bahdanau2014neural}, which achieves competitive translation performance, and a standard NMT Tensorflow implementation~\cite{luong17} is used. We leverage the Long-Short Term Memory (LSTM)~\cite{hochreiter1997long} as the RNN cell and consider three aspects that vary the design of the NMT model: (1) with or without the attention mechanism~\cite{bahdanau2014neural}, (2) different dropout rates, and (3) the number of recurrent layers.  To evaluate summarization performances, sentences that contain at least one action (43,582 out of 94,022 sentences satisfy the requirement) are randomly divided into a training set (23,582 sentences), a validation set (10,000 sentences), and a testing set (10,000 sentences); and each sentence corresponds to one or more reference summaries. NMT models are trained on the training set (batch size: 128, iterations: 100,000) using the Adam optimizer~\cite{kingma2014adam}, and the optimal training iterations are determined by the validation BLEU scores~\cite{papineni2002bleu}. Finally, models' performances are reported using the testing BLEU scores, as shown in Table.~\ref{nmt_results}.

\begin{table*}
\fontsize{7.5}{9.5}\selectfont
	\centering
	\begin{tabular}{>{\centering\arraybackslash}m{0.2cm} |>{\centering\arraybackslash}m{6cm} | >{\centering\arraybackslash}m{3.5cm} | >{\centering\arraybackslash}m{3.5cm} | >{\centering\arraybackslash}m{0.6cm}}
		\hline
		\rowcolor{gray}\textbf{Id} & \textbf{Sentence} & \textbf{NMT} & \textbf{References} & \textbf{BLEU}\\ \hline
		1 & create one more new layer just below the dress , set blending mode of the layer to `` soft light '' . & create one more new layer & \begin{tabular}[c]{@{}>{\centering\arraybackslash}m{3.5cm}@{}} create one more new layer\\ \hline set blending mode of the layer to`` soft light ''\end{tabular} & 1.0 \\ \hline
		2 & select your base image and duplicate ( control-j ) it . & duplicate layer & duplicate image & 0.84\\ \hline 
		3 & final image preview start working by creating a new document ( ctrl+n ) in adobe photoshop with the size 1900px by 1200px ( rgb color mode ) at a resolution of 72 pixels/inch . & creating new file & making new document & 0.76 \\ \hline
		4 & start by opening any image you want to work on . & opening photo of an image & open image & 0.67 \\ \hline
		5 & step 3 : load the photographic toning presets when the gradient picker appears , click on the small gear icon in the top right corner : clicking the gear icon . & load photographic toning presets and choose choose sepia $<$unk$>$ then gradient & click on the small gear icon & 0 \\ \hline
		6 & we want the white bands on the umbrella to be slightly translucent , so choose select $>$ color range , select the white stripes and apply these settings & controls the amount white area to blend over the stage details between better areas & select color & 0 \\ \hline
		7 & desaturate this cloud layer and use a big soft brush to erase the outer portion of the cloud , leave the bits as show below : change the blending mode to “ color burn ” and you will have the following effect : duplicate this cloud layer a few more times , position the duplicated layer around the edge of the cliff , adjust their blending mode as shown below : add a curves adjustment layer to it with mask as shown below : you will have the following effect : step 4 now let ’ s produce some lave and flaming effect to the image . & add blur effect & \begin{tabular}[c]{@{}>{\centering\arraybackslash}m{3.5cm}@{}} desaturate this cloud layer\\ \hline change to color burn\\ \hline duplicate this cloud layer \\ \hline adjust according to image shown below \\ \hline name it lava \end{tabular} & 0 \\ \hline
	\end{tabular}
	\caption{Usage summaries for sample sentences in the validation set. The instances are ordered by their BLEU scores descendingly. The summaries are generated from the 1-layer-att-dropout-0.5 NMT model. Each sentence may correspond to multiple ground-truth references.\label{qualitative_summary}}
\end{table*}

The quantitative results demonstrate that adding the attention layer and increasing the dropout rate significantly improve the summarization performance, but simply stacking more LSTM layers does not help. Compared to the machine translation task where the existing model~\cite{luong17} achieves the BLEU score close to 30, the best summarization model (BLEU: 21.53) is still sub-optimal.  To better understand errors, similar to Section \ref{sec:command_prediction}, we qualitatively analyze the outputs of the model \textbf{layer-1-att-dropout-0.5}. In Table.~\ref{qualitative_summary}, we present the summaries for sample sentences in the validation set, along with their groundtruth references. Our main findings are discussed as follows.
\begin{itemize}
	
	\item \textbf{Common and preliminary operations are easy to summarize.} These operations include creating or duplicating layers (\ref{qualitative_summary}-1, 2), and opening or loading documents (\ref{qualitative_summary}-3, 4). The model can easily pick up keywords, such as ``create'', ``duplicate'', ``layer'', ``document'' and ``file'', and generate clean summaries accordingly.
	
	
	\item \textbf{Long text and complex operations pose challenges to the summarization.} The summaries tend to be incomplete (\ref{qualitative_summary}-5, 6) or trivial (\ref{qualitative_summary}-7) when the input text is long (\ref{qualitative_summary}-7) or the operations are complex, e.g., adding diverse effects and involving multiple (\ref{qualitative_summary}-7) or long-tail (\ref{qualitative_summary}-5, 6) actions. 
\end{itemize}

To handle complex scenarios discussed above, summarization models need to paraphrase or accurately identify keywords and phrases from long text descriptions. In addition, similar to command predictions, understanding diverse effects that may be described is important for summarizing multiple or uncommon operations.

\section{Conclusions and Future work}

We formalized and collected annotations for CPK extractions, which are shown to pose new methodological research challenges. As discussed, CPK has potentials to power intelligent applications for information retrieval, personalized learning and autonomous design. Our future work includes further cleaning the collected annotations, building scalable knowledge extractors to populate and enrich CPK, and exploring tutorials with other formats, such as videos.

\bibliography{emnlp2018}
\bibliographystyle{acl_natbib_nourl}

\end{document}